\documentclass[amsthm,10pt]{elsart}
\usepackage{amsmath,amsthm,amsfonts,amssymb}
\usepackage[dvips]{graphicx}

\begin{document}

\begin{frontmatter}

\title{Couplings of hybrid operators to ground and excited states of bottomonia}

\author{Tommy Burch}, 
\author{Christian Ehmann}

\address{Institut f\"ur Theoretische Physik, Universit\"at 
Regensburg, D-93040 Regensburg, Germany}

\begin{abstract}
We analyze the overlap of local color-octet meson operators with the $\Upsilon$ and the $\eta_b$ and their low-lying excited states,
especially the first radial excitations. Our analysis is based on NRQCD and 
includes all terms up to order $v^4$. We use a variety of source and sink operators as a basis for the variational method, 
which enables us to clearly separate the mass eigenstates and hence to extract the desired amplitudes. 
The results show the usefulness of the variational method for determining couplings to excited hadronic states.
\end{abstract}


\end{frontmatter}

\section{Introduction}
The inner structure of hadrons is one of the most shrouded and difficult issues of modern physics. Due to the confining nature of strong interactions, perturbative QCD calculations cannot describe physics at hadronic energy scales from first principles. However, lattice QCD, as a nonperturbative approach, has succeeded in confirming and predicting properties of matter at these scales. Since {\it a priori} we are not well versed in the structure of hadronic states, we try to probe them on the lattice with suitable operators. In quantum field theory, the only condition for non-zero overlap of an eigenstate of the underlying Hamiltonian with a specific operator is that the operator has the same quantum numbers as the mass eigenstate. So all physical states, the ground state as well as all its excitations, are expected to couple non-trivially to currents which project out the corresponding quantum numbers. A recent paper by Liu and Luo \cite{Luo:2005zg}, addressed to spectroscopy of charmonia, suggests that there is an exception to this rule. In correlators, which they constructed from local color-octet (or hybrid) operators, no low-lying excitations are visible. Only the ground state and a much higher lying state, most probably a hybrid excitation, appear. We want to check this result for bottomonium systems in the pseudoscalar and vector channel.  Furthermore, we want to provide quantitative estimates for the couplings (another approach was recently used for excited pions \cite{McNeile:2006qy}).  To do so, we use the variational method, which is based on the construction of a cross correlator matrix. This approach enables us not only to investigate the spectrum, but more importantly, to obtain ratios of couplings of different local operators to a physical state. 

Of course, none of these operators, which merely serve as an expanding basis for the states, are physical in QCD, since mixing between them is possible. In the framework of NRQCD we can actually identify the terms in the Lagrangian, which are responsible for this, so called, configuration mixing. This has been studied in \cite{Burch:2003zf}, for instance. Since such configuration mixings are ever-present in physical hadronic systems, all states should have overlap with operators which have the same quantum numbers, regardless of whether they are hybrid in nature. Since hybrid operators contain elements of the field strength tensor, the couplings associated with them are divergent quantities. Rather than removing these divergences, it is our aim here to show that, at typical lattice scales, the hybrid couplings are finite, yet relatively small.

This paper is organized as follows. Section \ref{method} describes the variational method we use for extracting masses and couplings. Our implementation of the NRQCD framework to propagate the fermions is discussed in Section \ref{NRQCD}. Section \ref{configurations} gives an overview of the configurations on which we are running our simulation. The actual results are presented in Section \ref{results} and Section \ref{conclusions} contains concluding remarks.

A preliminary report of these results can be found in \cite{Ehmann:2006jd}. 

\section{Method}
\label{method}
To gain information on the couplings we rely on the variational method.
The starting point is the construction of a cross correlator matrix $C_{ij}$:
\begin{eqnarray}
 C_{ij}(t) & =& \langle 0 | \hat O_i(t) \hat{\overline O}_j(0) | 0 \rangle.
\end{eqnarray}
Then we can solve the generalized eigenvalue problem 
\begin{eqnarray}
 C_{ij}(t)\, \psi^n_j  & = & \lambda_n(t,t_0)\, C_{ij}(t_0) \psi^n_j ,
\end{eqnarray}
where $\lambda_n$ is the eigenvalue corresponding to the eigenvector $\vec\psi^n$.
Since, due to fluctuations, $C_{ij}$ is not exactly symmetric (although it is within errors), we symmetrize it by hand in order to make the diagonalization procedure more stable. 

The eigenvalues are then given by \cite{Mi85,LuWo90}
\begin{eqnarray}
\lambda_n(t,t_0) & = & A_n \, e^{-E_n(t-t_0)}[1+O(e^{-\Delta E_n(t-t_0)})],
\label{ev}
\end{eqnarray} 
where $E_n$ denotes the mass of the $n$th state and $\Delta E_n$ the mass difference to the state closest in mass to the state $n$. This correction is due to the use of a finite number of operators in the basis. However, for large enough values of $t$, we have a single mass eigenstate in each channel. Thus, the variational method enables us to clearly separate the ground state, many lower lying excitations, and even ghosts \cite{Burch:2005wd} (which do not play any role in this work due to the large mass of the quarks).

For our purposes here, even more important than the eigenvalues is the analysis of the eigenvectors. As a starting point consider the spectral decomposition of $C$:
\begin{equation}
C_{ij}(t) = \sum_{n=1}^{\infty} \frac{v_i^n {v_j^n}^*}{2E_n} e^{-E_nt}
\end{equation}
where $v_i^n = \langle 0| O_i | n \rangle \equiv \langle O_i | n \rangle$, that is the overlap of $i$th operator with the $n$th state. To extract this matrix element consider the right-multiplication of the correlator matrix with one of the eigenvectors:
\begin{equation}
\label{recon}
\sum_j C_{ij}(t) \, \psi^n_j = \frac{1}{2E_n}v_i^n a^n e^{-E_nt} [1+O(e^{-\Delta E_nt})] \; \stackrel{t\gg t_0}{\longrightarrow} \; \frac{1}{2E_n}v_i^n a^n e^{-E_nt},
\end{equation}
where $a^n=\sum_i v^{n*}_i \psi^n_i$.
For large enough times a single mass eigenstate is projected out of the sum (see Eq. (\ref{ev})). 
We can now left-multiply with the same eigenvector to obtain 
\begin{equation} 
\sum_{i,j}  \psi^{n*}_i \, C_{ij}(t) \, \psi^n_j \approx \frac{1}{2E_n} |a^n|^2 e^{-E_nt}. 
\end{equation}
In the event that the correlator matrix is purely real (or purely imaginary) we can calculate the $a^n$ and divide them out from Eq.(\ref{recon}) to obtain the couplings $v_i^n$.
Since we are only interested in ratios of couplings, we proceed with Eq.(\ref{recon}).
To cancel the operator independent amplitude, the energy prefactor and the exponential we construct a ratio of the following kind:
\begin{equation}
\label{recons}
\frac{\sum_j C_{ij}(t)\psi^n_j}{\sum_l C_{kl}(t)\psi^n_l} \, 
\stackrel{t\gg t_0}{\longrightarrow} \,
\frac{\langle O_i|n\rangle}{\langle O_k|n\rangle} = 
\frac{Z^{-1}_i{\langle O_i|n\rangle}_{ren}}{Z^{-1}_k{\langle O_k|n\rangle}_{ren}},
\end{equation}
where $Z_i$ is the renormalization constant for the $i$th operator and ${\langle \ldots \rangle}_{ren}$ denotes the renormalized quantity. 

It is important to note that in using a finite, non-orthogonal basis of operators, the individual overlaps may not be well defined. It is vital that one checks, for example, that the above ratio of local couplings does not change when further operators are added to the basis.

The success of the variational method strongly depends on the choice of operators one includes in the basis. Unfortunately, there will in any case only be a finite number of operators available, which can just span a subspace of the Hilbert space of states. Therefore, one tries to use operators which have small relative overlap in order to explore a large area of the space the meson wavefunctions live in.
Since we want to investigate the $\eta_b$ and the $\Upsilon$, we use pseudoscalar and vector currents. For both we have a ``normal'' and a ``hybrid'' version. Table \ref{optab} gives an overview of the local operators we use. It additionally contains operators coupling to states which we only address in the spectrum section. The P-wave states are moreover needed to set the scale. In order to assemble our basis with more linearly independent operators, we additionally smear the quark and the antiquark field independently with two different smearing levels $n_{narrow}=16$ and $n_{wide}=48$ using gauge invariant Jacobi smearing with $\kappa=0.2$. 

In total, twelve different operators, each at the source and the sink, are available for constructing the cross correlator matrix. 

\vspace{1cm}
\begin{table}[!h!]
\begin{center}
 \begin{tabular}{|c|c|c|c|}
 \hline
 state & $J^{PC}$ & normal operator & hybrid operator \\
 \hline
 $\eta_b$ & $0^{-+}$ & $\chi^{\dagger}\phi$ & $\chi^{\dagger}\sigma_i B_i \phi$  \\
 $\Upsilon$ & $1^{--}$ & $\chi^{\dagger}\sigma_i\phi$ & $\chi^{\dagger}B_i \phi$  \\
 $\chi_{b0}$ & $0^{++}$ & $\chi^{\dagger}\sigma_iD_i\phi$ & - \\
 $\chi_{b1}$ & $1^{++}$ & $\chi^{\dagger}\epsilon_{ijk}\sigma_jD_k\phi$ & - \\
 $\chi_{b2}$ & $2^{++}$ & $\chi^{\dagger}(\sigma_iD_j+\sigma_jD_i-\frac{2}{3}\delta_{ij}\sigma_kD_k)\phi$ & - \\
 $h_b$ & $1^{+-}$ & $\chi^{\dagger} D_i \phi$ & - \\
 D-wave & $2^{-+}$ & $\chi^{\dagger}(D_iD_j+D_jD_i-\frac{2}{3}\delta_{ij}D_kD_k)\phi$ & - \\ 
 exotic & $1^{-+}$ & - & $\chi^{\dagger}\epsilon_{ijk}\sigma_jB_k\phi$  \\
\hline 
\end{tabular}
\caption{\label{optab} Overview of the used operators $\hat O_i$. The meaning of $\phi$ and $\chi$ will be defined in the subsequent section.} 
\end{center}
\end{table}

\section{NRQCD}
\label{NRQCD}
The calculation of the propagators is performed in the framework of NRQCD (see \cite{Lepage:1992tx}), which is perfectly suitable for our bottomonium systems, where the quarks move with small relative velocities. We include all terms up to $O(v^4)$ in our NRQCD Lagrangian, where $v$ is the velocity of a quark, according to the power counting in \cite{Lepage:1992tx}. Since we are working in a nonrelativistic approximation, it is convenient to separate the Dirac four spinor $\psi$ into two sets of Pauli spinors, $\phi$ for the quark and $\chi$ for the antiquark. Furthermore, the propagation of the fermions can be described by the quantum mechanical evolution operator for imaginary time $e^{-Ht}$. By expanding this operator we obtain for the propagation:
\begin{equation}
\label{evolve}
\phi(\mathbf{y},t+a)=\left(1-\frac{aH_0(t+a)}{2n}\right)^nU_4^\dagger(t)\left(1-\frac{aH_0(t)}{2n}\right)^n\left(1-a\delta H(t)\right)\phi(\mathbf{x},t),
\end{equation}
where $H_0$ is
\begin{eqnarray}
H_0 & = & -\frac{\tilde{\Delta}}{2M}-\frac{a}{4n}\frac{\tilde{\Delta}^2}{4M^2}
\end{eqnarray}
and $\delta H$ is
\begin{eqnarray}
\label{rel_corr}
\delta H & = & -\frac{1}{8M^3}\tilde{\Delta}^2 \\ \nonumber
& & +\frac{ig}{8M^2}(\nabla\cdot\mathbf{E}-\mathbf{E}\cdot\nabla)-\frac{g}{8M^2}\mathbf{\sigma}\cdot(\tilde{\nabla}\times\mathbf{E}-\mathbf{E}\times\tilde{\nabla}) \\ \nonumber
& & -\frac{g}{2M}\mathbf{\sigma}\cdot\mathbf{B}.
\end{eqnarray}
The tildes denote improved versions of the corresponding derivatives. We use $n=2$, which is more than sufficient in our case. 
$\mathbf{B}$ and $\mathbf{E}$ are the magnetic and electric fields created via the usual clover formulation. The last two terms of (\ref{rel_corr}) are responsible for the configuration mixing mentioned above; however, it turns out that the $\mathbf{\sigma}\cdot\mathbf{B}$ term produces a much larger effect.

To correct for tadpole contributions we divide each link by a factor $u_0$ which is given by the fourth root of the average plaquette \cite{Lepage:1992xa}. The quark mass for our simulation is determined from finite momentum correlators for the $\Upsilon$ by tuning the kinetic mass extracted from the non-relativistic energy-momentum dependence to the experimental mass of the $\Upsilon$.

\section{Configurations}
\label{configurations}
We are working on configurations provided by the MILC-collaboration \cite{Aubin:2004wf}. They were generated using improved staggered fermions and the L\"uscher-Weisz gauge action. Table {\ref{configs} shows the parameters of the lattices used. For the lattice spacing, there are two values given. The first one comes from the analysis of the spin-averaged $\Upsilon$ 1P-1S splitting, the second one is given by the MILC-collaboration, where they used the improved Sommer parameter $r_1$ to set the scale. Note that for the quenched lattice we are slightly above the values from the MILC-Collaboration. This is most likely due to the wrong curvature of the $q\bar{q}$ potential in quenched simulations, thus influencing the P- and the S-waves differently.

To obtain the physical $b$-quark mass at the corresponding scales, we inter-/extrapolate to the experimental $\Upsilon$ mass. However, both for the spectra and the ratios of the couplings, we find a very weak dependence upon the quark mass parameter.

\vspace{0.5cm}
\begin{table}[h!]
\begin{center}
 \begin{tabular}{|c|c|c|c|c|c|c|c|}
 \hline
$\beta$ & $a^{-1}[\mathrm{MeV}]$ & volume & $N_f$ & $am_{sea}$ & $am_{b}$ & $m_b^{phys}[\mathrm{MeV}]$ & \# of configs. \\
\hline
8.40 & 2378(7)/2279 & $28^3\times96$ & 0 & $\infty$ & 1.7, 1.8 & 4114(72) & 160 \\
7.09 & 2097(9)/2252& $28^3\times96$ & 2+1 & 0.0062/0.031 & 1.7, 1.8 & 4026(65) & 210 \\
6.76 & 1495(6)/1587& $20^3\times64$ & 2+1 & 0.01/0.05 & 2.4, 2.5 & 4111(48) & 410 \\
\hline
\end{tabular}
\end{center}
\caption{\label{configs} The three different lattices we use in our simulations together with the corresponding parameters. The first value for the inverse lattice spacing comes from the 1P-1S splitting, the second one is given in \cite{Aubin:2004wf}.}
\end{table}

\section{Results}
\label{results}
\subsection{Spectra}
At first we want to present the results for the masses in the considered channels. Several efforts have already been made to investigate the spectrum of bottomonia (\cite{Davies:1994mp}, \cite{Davies:1998im}, \cite{Juge:1999ie}, \cite{Davies:2003ik}).
Table \ref{spectab} gives some of the S-wave splittings for the fine quenched and dynamical lattices. Figure\ \ref{specfig} shows the plot of the absolute masses. The fits are one exponential fits to the corresponding correlators of the states, using the full covariance matrix. For the pseudoscalar and the vector channel we use the variational method in the basis Nll, Nln, Nnn, Nww, Hll \footnote{The capital letter denotes the type of the operator: N=normal, H=hybrid; the two small ones give the smearing of the quark and the antiquark respectively: l=local, n=narrow, w=wide.} and thus are able to obtain masses of their excited states reliably, the masses of the other states are extracted from single correlators. The absolute mass offset is fixed by the value of the $\Upsilon$. For both types of configurations the quark mass parameter was set to $am_b=1.7$. It is worth noting that the experimental input for the $\eta_b$ comes from a single event, so it is not very reliable.

\vspace{-.3cm}

\begin{figure}[h!]
\begin{center}
\resizebox{380pt}{!}{\includegraphics[clip,angle=270]{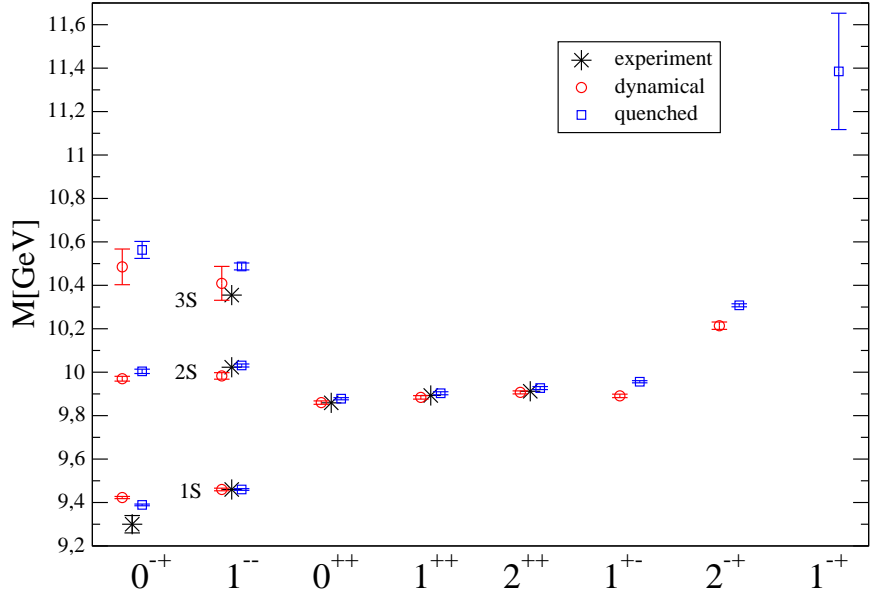}}
\end{center}
\caption{Spectra of the available states, obtained by fitting the corresponding correlators, for both quenched($\beta=8.40$) and dynamical configurations($\beta=7.09$).}
 \label{specfig}
\end{figure}

\begin{figure}[h!]
\begin{center}
\resizebox{350pt}{!}{\includegraphics[clip,angle=270]{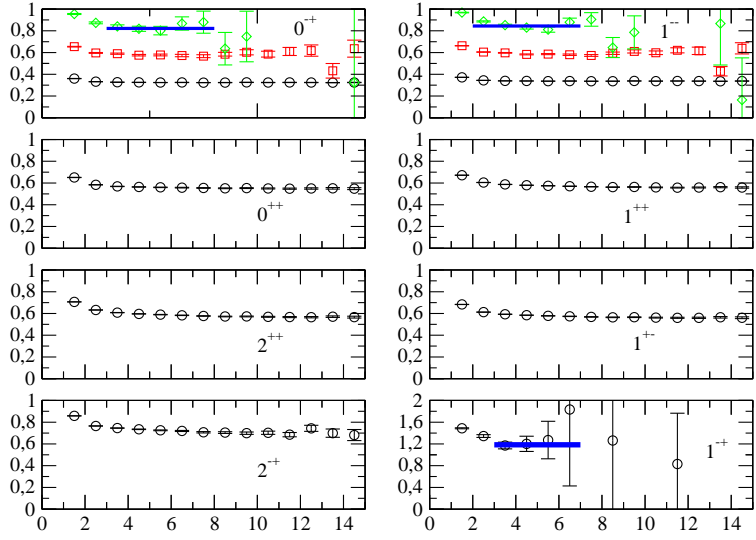}}
\end{center}
\caption{Effective mass plots of the available states from the quenched lattice with $\beta=8.40$. The results for $0^{-+}$ and $1^{--}$ come from the variational method, the other effective masses have been obtained from single correlators. The blue lines indicate the fitted plateaus. The fitting errors can be told from their width. Note the different scale for the exotic $1^{-+}$.}
 \label{effmassquen}
\end{figure}

\vspace{-1cm}

\begin{figure}[h!]
\begin{center}
\resizebox{350pt}{!}{\includegraphics[clip,angle=270]{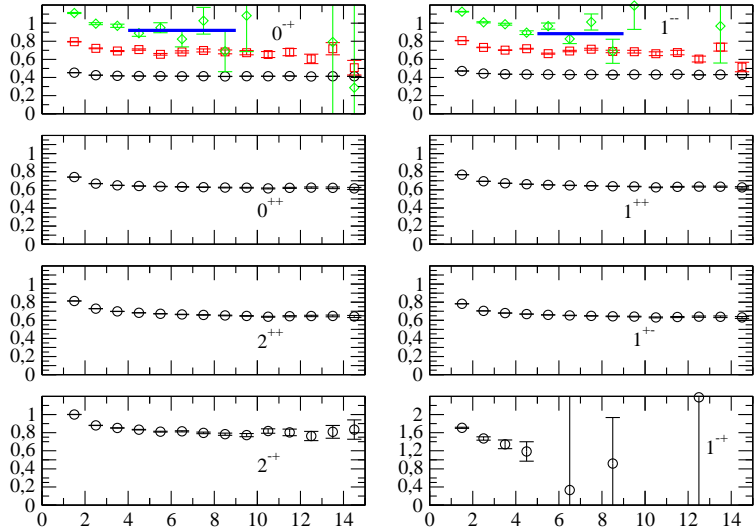}}
\end{center}
\caption{Same as in Fig.\ \ref{effmassquen} but for the dynamical lattice with $\beta=7.09$.}
\end{figure}

\begin{table}[h!]
\begin{center}
\begin{tabular}{|c|c|c|c|c|c|c|}
\hline
 & $0^{-+}(1S)$ & $0^{-+}(2S)$ & $0^{-+}(3S)$ & $1^{--}(1S)$ & $1^{--}(2S)$ & $1^{--}(3S)$ \\
 \hline
 \hline
\bf{experimental} & 9.300(28) & - & - & 9.460 & 10.023 & 10.355\\
 \hline
\bf{quenched} &  9.389(4) & 10.004(10) & 10.563(39) & 9.460(4) & 10.031(7) & 10.487(16) \\
$\chi^2/dof$ & 2.49/5 & 7.42/4 & 2.50/4 & 6.76/5 & 2.36/4 & 5.64/4 \\
fit range & 4-10 & 3-8 & 3-8 & 4-10 & 3-8 & 2-7 \\
 \hline
\bf{dynamical} & 9.423(5) & 9.970(11) & 10.485(82) & 9.460(6) & 9.983(15) & 10.409(78) \\
$\chi^2/dof$ & 4.56/5 & 6.56/5 & 2.43/4 & 2.65/4 & 1.71/3 & 1.27/3 \\
fit range & 4-10 & 4-10 & 4-9 & 10-15 & 7-11 & 5-9 \\
\hline
\hline
 & $0^{++}$ & $1^{++}$ & $2^{++}$ & $1^{+-}$ & $2^{-+}$ & $1^{-+}$ \\
\hline
\hline
\bf{experimental} & 9.859 & 9.893 & 9.912 & - & - & -\\
\hline
\bf{quenched} & 9.878(5) & 9.903(5) & 9.927(7) & 9.956(5) & 10.308(7) & 11.385(268) \\
$\chi^2/dof$ & 2.79/5 & 4.52/4 & 2.44/2 & 5.89/5 & 8.03/7 & 0.74/2 \\
fit range & 10-17 & 12-18 & 16-19 & 10-16 & 7-15 & 3-7  \\
\hline
\bf{dynamical} & 9.860(8) & 9.884(8) & 9.907(7) & 9.891(8) & 10.214(17) & -  \\
$\chi^2/dof$ &  3.15/4 & 5.09/4 & 3.04/4 & 3.02/3 & 4.63/4 & - \\
fit range &  11-16 & 11-16 & 11-6 & 11-15 & 9-14 & -\\
\hline
\end{tabular}
\caption{Fitting results for the masses in GeV for both quenched($\beta=8.40$) and dynamical configurations($\beta=7.09$).}
 \label{spectab}
\end{center}
\end{table}

\clearpage

\subsection{Starting basis}
Local operators are well defined in the context of quantum field theory and therefore the ratio of their couplings is of particular interest. Furthermore, this will provide us with a clear answer if radial\footnote{The low-lying excited states of bottomonia of definite $J^{PC}$ are ``mostly'' radial excitations. Of course, configuration mixings will ensure that some small admixtures of other configurations (gluonic, orbital, etc.) are present. However, in the following we use ``radially excited'' and ``excited'' interchangeably.} excitations have overlap with local hybrid operators.
To access these ratios we proceed to the analysis of the eigenvectors of the cross correlator matrix. As described above we construct the correlator matrix with a variety of source and sink operators and diagonalize it. By this procedure we hope to clearly separate the mass eigenstates and use the eigenvectors to shed light on their overlap with the included operators. 

We start our analysis for the $\Upsilon$, which is probably the experimentally more interesting state, on the dynamical lattice with $\beta=7.09$ and a quark mass parameter of $am_b=1.7$. Later on we will also investigate the other lattices with different parameters to see what changes arise. 

The smallest basis, including a hybrid operator, which reveals reasonable results, is Nll(1), Nnn(2), Hll(3). The numbers in brackets are given just for the sake of clarity in the coupling ratio plots. The effective masses of the eigenstates are shown in Fig.\ \ref{ev3}. The ground state and the first radial excitation are clearly visible.

To check if our diagonalization worked correctly, we can reconstruct the eigenvalue by multiplying the eigenvector with the cross correlator matrix $C_{ij}(t)\psi^n_j$. The indices $i,j$ label the operators and $n$ the states. This product should have the same exponential behavior, i.\,e., the same mass, as the corresponding eigenvalue. Of course, they may differ in their amplitude since $C_{ij}(t)\psi^n_j$ gives the eigenvalue times some overlap factor. But much more important is the fact that these overlap factors ultimately provide us with the ratios of the amplitudes. To give a quantitative estimate for the ratio, we have to plot the ratio with respect to time. When the ratio has plateaued, we can be sure that higher excited states do not play a role any more. 

Figure\ \ref{ratio3} shows the ratio of the overlaps of the local hybrid and the local normal operator for the ground state and the first radial excitation. We clearly find that radial excitations have non-zero overlap with local hybrid operators. This is obvious since the radial excitations have the same quantum numbers the hybrid operator couples to. In both cases the coupling of the local normal operator to the state is about 90 times larger than the coupling of the local hybrid operator. Of course, these numbers still require renormalization.
However, this is beyond the scope of this work since we are only interested in showing that the couplings to hybrid operators are finite, yet small for typical lattice scales, and that the variational method can be used to extract these couplings. 
By constructing a ratio of these ratios we define the $R$ value:
\begin{equation}
\label{ROR}
R^{nn'}_{ik} = \frac{{\langle O_i|n\rangle}_{ren}}{{\langle O_k|n\rangle}_{ren}}\left/\frac{{\langle O_i|n'\rangle}_{ren}}{{\langle O_k|n'\rangle}_{ren}} \right.,
\end{equation}
where, of course, $i$ and $k$ are not summed over.
According to Eq. (\ref{recons}) $R$ can be expressed in terms of the cross correlator matrix and its eigenvectors:
\begin{equation}
R^{nn'}_{ik} = \frac{\sum_j C_{ij}(t)\psi^n_j}{\sum_l C_{kl}(t)\psi^n_l}\left/\frac{\sum_r C_{ir}(t)\psi^{n'}_r}{\sum_s C_{ks}(t)\psi^{n'}_s} \right.
\end{equation}
The renormalization constants which only depend on the current cancel and a further cutoff dependent factor which may come from disagreeing mass dimensions of the operators cancels too. If one of the operators is a hybrid one, one would also expect a scale dependence due to its nonultralocal nature, more precisely $F_{\mu\nu}$ appearing in the hybrid operator is constructed from a 2x2 clover. We assume the factorization of this dependence and thus Eq. (\ref{ROR}) provides a scale independent physical quantity. We will actually show this in Section \ref{scaledep}.
Therefore, in order to remove any cutoff dependence, we utilize Eq. (\ref{ROR}) and obtain as a physical result that the ratio is about equal for the ground and the first excited state (see Fig.\ \ref{RORfig1}). 
The fit to a constant from $t=3$ to $t=10$ yields $R^{1,2}_{3,1} = 1.0464(63)$, where the error comes from single elimination jackknifing. 

\vspace{2cm}

\begin{figure}[h!]
\begin{center}
\resizebox{350pt}{!}{\includegraphics[clip,angle=270]{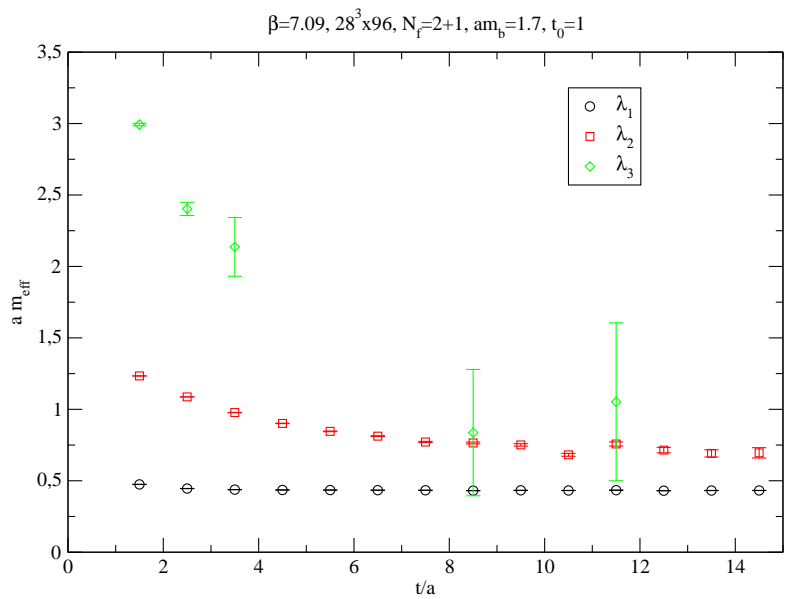}}
\end{center}
\caption{Effective masses of the three eigenvalues in the basis Nll(1), Nnn(2), Hll(3) for the dynamical lattice with $\beta=7.09$ and $am_b=1.7$. $t_0$ is the normalization timeslice.}
\label{ev3}
\end{figure}

\clearpage

\begin{figure}[h!]
\begin{center}
\resizebox{330pt}{!}{\includegraphics[clip,angle=270]{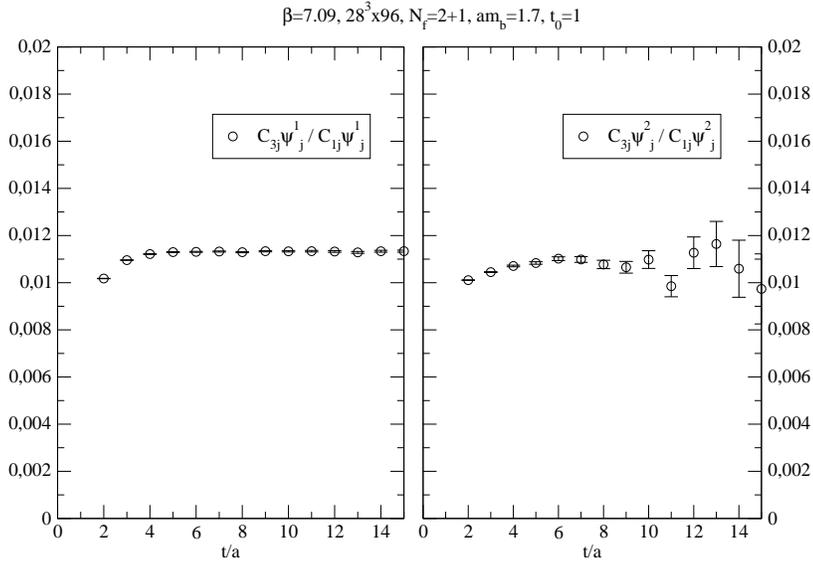}}
\end{center}
\caption{The ratio of the couplings of the two local operators to the ground state is shown in the left plot, the same for the first excitation is in the right plot.}
\label{ratio3}
\end{figure}

\vspace{2cm}

\begin{figure}[h!]
\begin{center}
\resizebox{300pt}{!}{\includegraphics[clip]{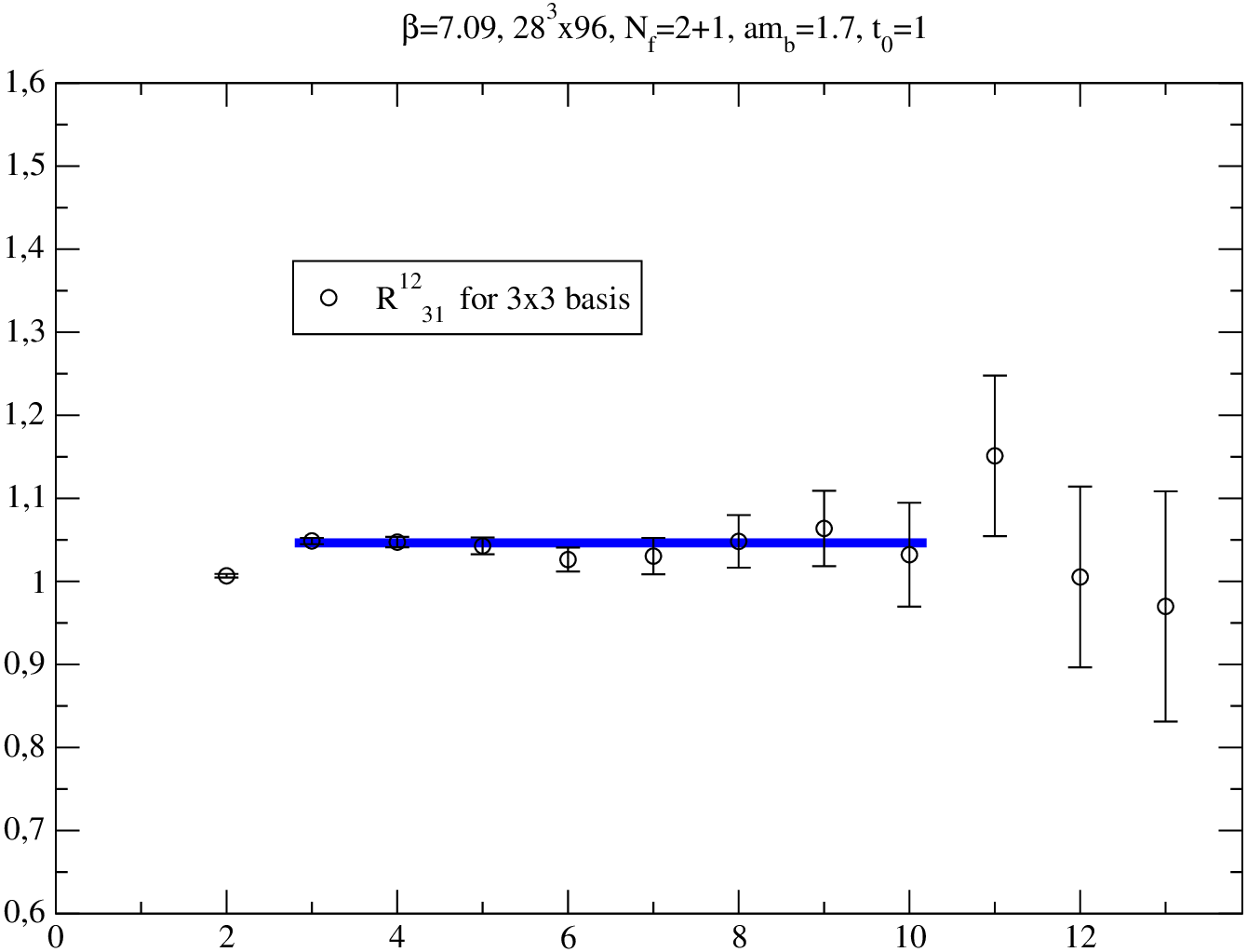}}
\end{center}
\caption{Plot of $R^{1,2}_{3,1}=\frac{{\langle O_3|1\rangle}_{ren}}{{\langle O_1|1\rangle}_{ren}}\left/\frac{{\langle O_3|2\rangle}_{ren}}{{\langle O_1|2\rangle}_{ren}} \right. $ .The blue line shows the result for fitting to a constant. } 
\label{RORfig1}
\end{figure}

\clearpage

\subsection{Enlarged basis}
\label{enlhybbas}
In order to obtain better signals from our correlators and to show a general feature of our operators, we increase the number of normal operators in the basis. The enlarged basis is built up from Nll(1), Nln(2), Nnn(3), Nww(4), Hll(5). Figure\ \ref{ev4} shows the effective masses of the eigenvalues. For this choice of basis we manage to obtain an acceptable signal even for the fourth state. The ratios of the couplings for the ground state and the first radial excitation Fig.\ \ref{ratio4}. Figure\ \ref{ratio6} shows the ratios for the third and fourth state, two further excitations which have been skipped in the previous basis. Again the results are close to those for the ground state and the first radial excitation. The value of $R$ evaluated in this basis for the ground and first excited states is shown in Fig.\ \ref{ROR2fig}. The fit to a constant in the range $t=3$ to $t=10$ gives $R^{1,2}_{5,1} = 1.0348(87)$, confirming the result from the previous basis. 

A property of local operators reveals itself when we look again at the ratio of amplitudes. A comparison of the ground state and the first radial excitation ratio in two different bases is shown in Fig.\ \ref{ratio5}. The explicit forms of the bases are: Basis 1 - Nll(1), Nln(2), Nnn(3), Nww(4), Hll(5), basis 2 - Nll(1), Nln(2), Nnn(3), Hll(4), Hln(5), Hnn(6). Even though, we change our basis by substituting smeared normal operators by other smeared ones, the ratios of the couplings of the local operators remain the same for these states. This implies that local operators are ``approximately'' orthogonal to smeared ones (i.\,e., they have very small spatial overlap). If they would overlap significantly, adding a smeared operator to the basis would change the ratio, because the smeared operator could ``steal'' some contribution of the projection of the physical state onto the local operator. When looking at ratios of couplings of two smeared operators, we see that they change significantly when another smeared operator is added to, or removed from, the basis. That means that the smeared operators we use are not orthogonal to each other. Since the coupling ratio of the local operators stays the same in every arbitrary basis, as we found, we can also exclude that the smeared operators occasionally steal exactly the same contribution from the local normal and the local hybrid operator in such a way that the ratio remains the same. The result about the orthogonality of local and smeared operators is very plausible. The spatial width of the local operators is near zero and they have finite height. So their convolution with the Jacobi smeared operators should be quite small, as we found. The constancy of the ratio under the change of basis is an important justification for our conclusions.

It is worth noting that including further smeared operators in the basis does not help to improve the outcome; quite the contrary, they enhance the overlap with higher excited states or they contribute more noise than new information and thereby disrupt the signals.

\vspace{2cm}
\begin{figure}[h!]
\begin{center}
\resizebox{330pt}{!}{\includegraphics[clip,angle=270]{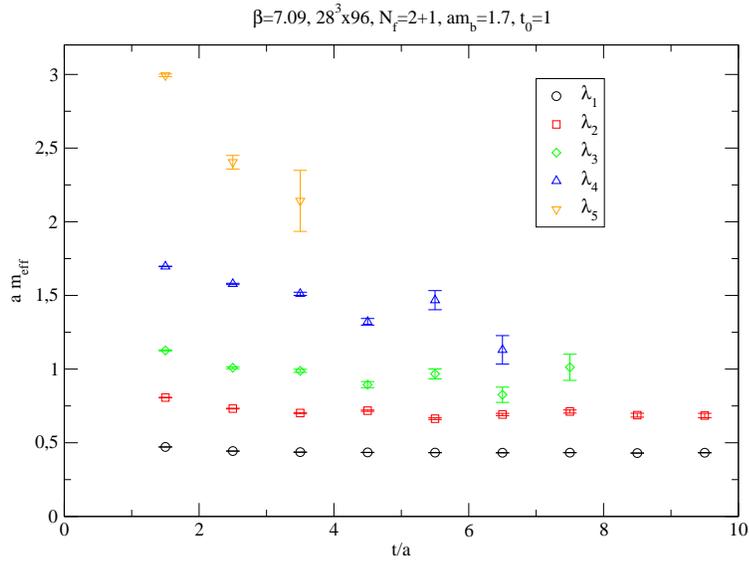}}
\end{center}
\caption{Effective masses of the five eigenvalues in the basis Nll(1), Nln(2), Nnn(3), Nww(4), Hll(5) for the dynamical lattice with $\beta=7.09$ and $am_b=1.7$.}
\label{ev4}
\end{figure}

\vspace{1cm}

\begin{figure}[h!]
\begin{center}
\resizebox{300pt}{!}{\includegraphics[clip,angle=0]{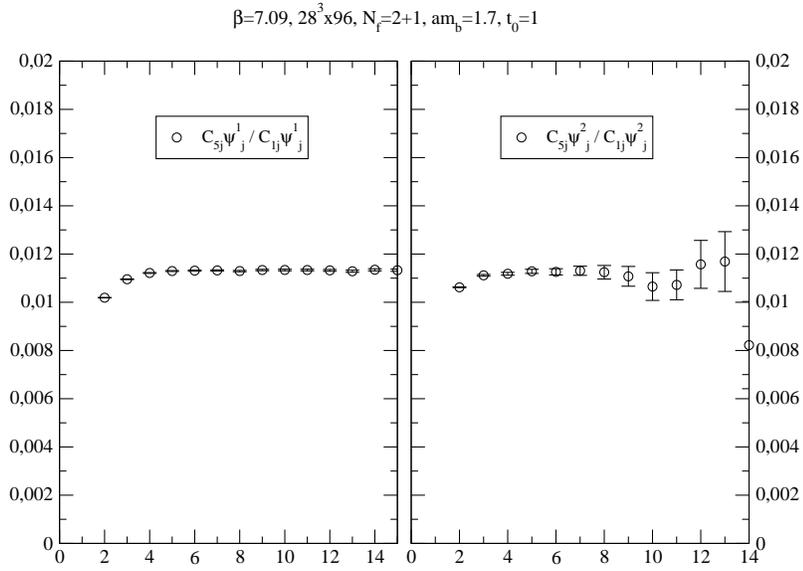}}
\end{center}
\caption{The ratio of the couplings of the two local operators to the ground state and to the first radial excitation. Both in the new 5$\times$5 basis.}
\label{ratio4}
\end{figure}

\begin{figure}[h!]
\begin{center}
\resizebox{330pt}{!}{\includegraphics[clip,angle=270]{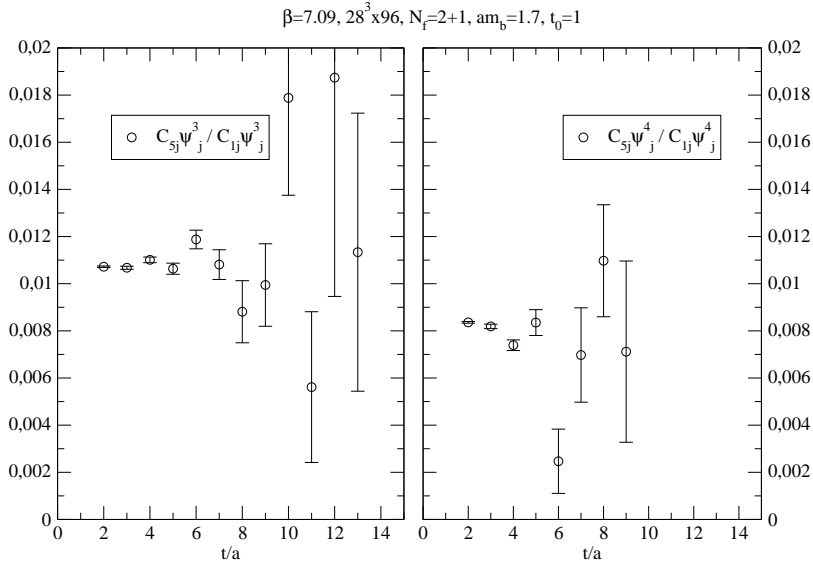}}
\end{center}
\caption{The ratio of the couplings of the two local operators to the third state is shown in the left plot, the one for the fourth state is in the right plot. Both in the new 5$\times$5 basis.}
\label{ratio6}
\end{figure}

\begin{figure}[h!]
\begin{center}
\resizebox{300pt}{!}{\includegraphics[clip]{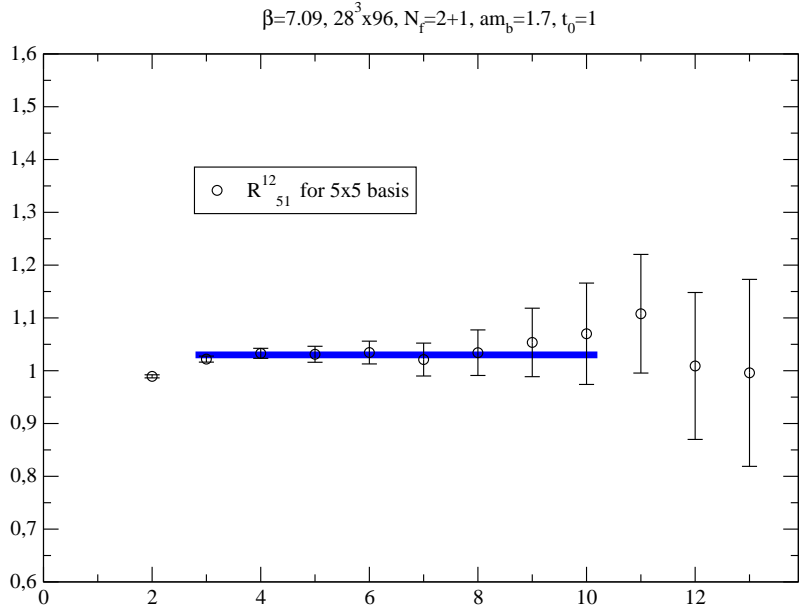}}
\end{center}
\caption{Plot of $R^{1,2}_{5,1} = \frac{{\langle O_5|1\rangle}_{ren}}{{\langle O_1|1\rangle}_{ren}}\left/\frac{{\langle O_5|2\rangle}_{ren}}{{\langle O_1|2\rangle}_{ren}} \right. $. The blue line shows the result for fitting to a constant. } 
\label{ROR2fig}
\end{figure}

\begin{figure}[h!]
\begin{center}
\resizebox{330pt}{!}{\includegraphics[clip,angle=270]{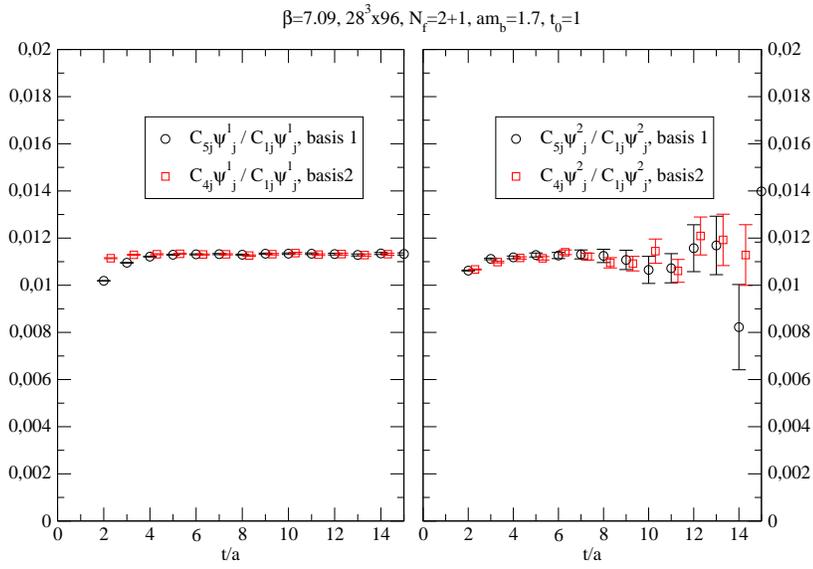}}
\end{center}
\caption{The ratio of the couplings of the two local operators to the ground state is shown in the left plot, the one for the first excitation is in the right plot, both for two different sets of operators [basis 1 is Nll(1), Nln(2), Nnn(3), Nww(4), Hll(5) and basis 2 Nll(1), Nln(2), Nnn(3), Hll(4), Hln(5), Hnn(6)]. The datapoints for the second basis are shifted by 0.3 for the sake of clarity.}
\label{ratio5}
\end{figure}

\clearpage
\subsection{Scale dependence/quenching effects}
\label{scaledep}
To verify that $R$ defined in Eq. (\ref{ROR}) really is scale independent, as we claim, we do the same calculation in the previous 5x5 basis for a coarser lattice ($a^{-1}=1497$ MeV compared to $a^{-1}=2101$ MeV in the previous calculation) with $am_b=2.4$. The results are shown in Fig.\ \ref{ratio10} and Fig.\ \ref{ror_quot}. From Fig.\ \ref{ratio10} follows a clear scale dependence for the individual couplings. However, Fig.\ \ref{ror_quot} certifies that this dependence vanishes for $R$. There $R^{1,2}_{5,1}$ fitted to a constant in time at three different scales, two from dynamical and one from quenched lattices, is plotted. Just considering the case with sea quarks included we see that the two points are consistent within the errors, proving that any scale dependence is very small.
To guide the eye we performed a fit to a constant with the two dynamical points, yielding 1.00784(836), which is consistent with one.
The fact that the quenched result is relatively low shows the different effect of virtual quark-antiquark pairs for the ground and the first excited state. However, the hybrid coupling is also seen to be small on the quenched lattices.

\vspace{1cm}
\begin{figure}[h!]
\begin{center}
\resizebox{330pt}{!}{\includegraphics[clip,angle=270]{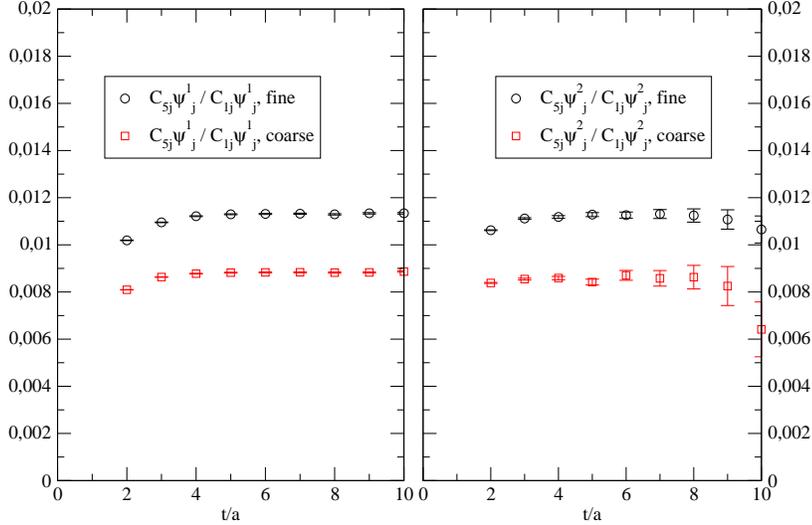}}
\end{center}
\caption{The ratio of the couplings of the two local operators to the ground state is shown in the left plot, the one for the first excitation in the right plot. Each plot is for two different lattices with $a^{-1}=2101$ MeV (fine) and $a^{-1}=1497$ MeV (coarse).}
\label{ratio10}
\end{figure}

\vspace{1cm}
\begin{figure}[h!]
\begin{center}
\resizebox{330pt}{!}{\includegraphics[clip,angle=0]{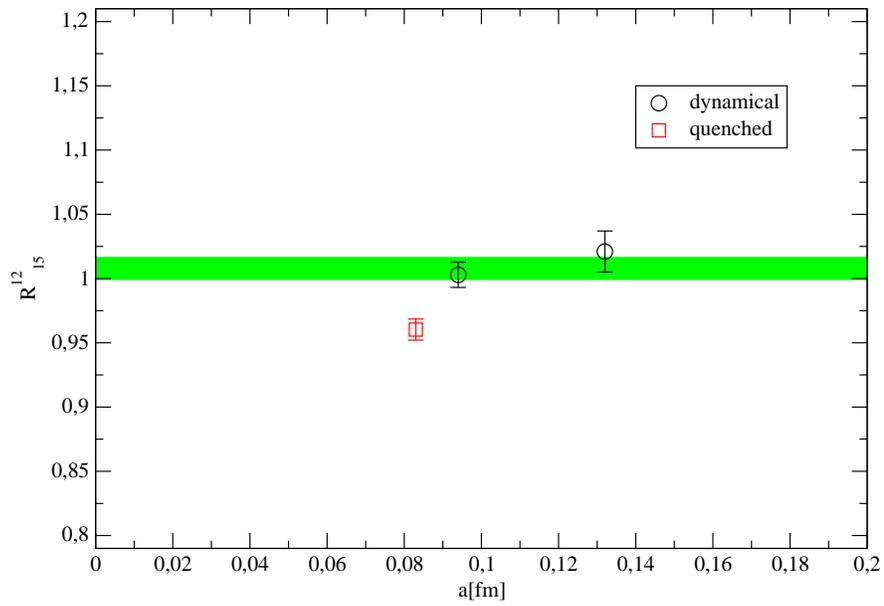}}
\end{center}
\caption{Plot of $R^{1,2}_{5,1}$. The green line shows the result for fitting the dynamical points to a constant. The fitting error can be told from its width.}
\label{ror_quot}
\end{figure}

\clearpage

\subsection{Further results}
This section briefly covers the results for the $\eta_b$, which is, from an experimental viewpoint, less accessible, since its ground state has been observed in only one type of experiment and none of its excitations have been detected up to now.

Figure\ \ref{ev9} shows the effective masses of the eigenvalues in the basis Nll(1), Nln(2), Nnn(3), Nww(4), Hll(5). The local coupling ratios are plotted in Fig.\ \ref{ratio13}. Typically, the signals for pseudoscalars are slightly better than the ones for vectors. 
The increase of the ratio of the local couplings by about a factor three can be traced back to the fact that in the $0^{-+}$ hybrid operator all three components of the $B$-field are included.

\vspace{1cm}
\begin{figure}[h!]
\begin{center}
\resizebox{330pt}{!}{\includegraphics[clip,angle=270]{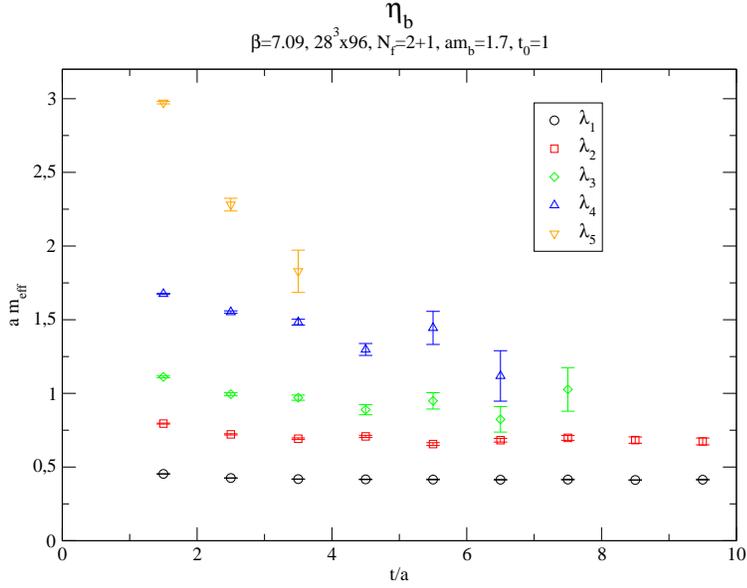}}
\end{center}
\caption{Effective masses of the five eigenvalues in the basis Nll(1), Nln(2), Nnn(3), Nww(4), Hll(5) for the pseudoscalar on the dynamical lattice with $\beta=7.09$ and $am_b=1.7$. }
\label{ev9}
\end{figure}

\begin{figure}[h!]
\begin{center}
\resizebox{300pt}{!}{\includegraphics[clip,angle=270]{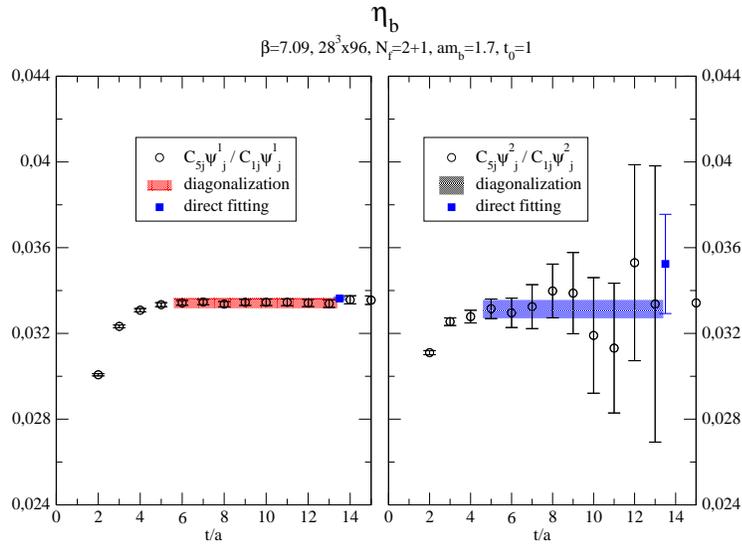}}
\end{center}
\caption{The ratio of the couplings of the two local operators to the ground state is shown in the left plot, the one to the first excitation in the right plot. Both for the pseudoscalar state on the dynamical lattices. Additionally the results for fitting the ratios to a constant with and without diagonalization are shown. The fit range for the direct fit of the ($3\times3$) correlator matrix is $t=13-18$ for both states.}
\label{ratio13}
\end{figure}

\clearpage

As a crosscheck we also determined the couplings of the two local operators to the two lowest lying states by directly fitting a three by three correlator matrix. To do so, we build up a matrix of correlators with three different operators, namely Nll(1), Nnn(2), Hll(3). Then we fit the matrix to the functional form 
\begin{equation}
C_{ij}(t)=A^0_iA^0_je^{-aE_0t}+A^1_iA^1_je^{-aE_1t}, \quad i,j=1,2,3 ,
\end{equation} 
where $A^0_i$ is the coupling of the ground state to the $i$th operator and $A^1_i$ is the coupling of the first excited state to the $i$th operator. We find that fitting the above form with a larger basis of operators and with more masses becomes highly problematic.

The results for the ratios of couplings and the corresponding $R$ value obtained by direct fitting together with the ones after diagonalization in the basis Nll(1), Nln(2), Nnn(3), Nww(4), Hll(5) are given in Table \ref{cmfit}, both for the vector and the pseudoscalar channel (see also Fig.\ \ref{ratio13}). The time fit range to determine the couplings in the case of direct fitting is $t=13$ to 18 for all three quantities, in the case of the fit after diagonalization $t=6$ to 13 for the ground state and $t=5$ to 13 for the excited state and $R$. The range for direct fitting multiplied by the nine different operator combinations yields 54 degrees of freedom to determine eight parameters, namely six couplings and two masses. The errors in both cases were calculated by single elimination jackknifing. The numbers suggest that the direct fitting method is slightly better for the ground state, but much worse for the excited one. One may criticize such a comparison by pointing out that the direct fitting takes place with only three operators in the basis, while the diagonalization works with five. But actually this is the main virtue of the variational approach. With the diagonalization, one is able to work with more operators, thus putting in more information. In turn, this provides eigenvalues which more readily correlate single mass states. One can then start fitting (single exponentials) much earlier, thereby avoiding the noisier areas of the time range. However, we note that the outcomes of both methods show reasonable agreement.

\vspace{1cm}

\begin{table}[h!]
\begin{center}
\begin{tabular}{|c|c|c|c|c|}
\hline
  & \multicolumn{2}{|c|}{pseudoscalar} &  \multicolumn{2}{|c|}{vector} \\
 \cline{2-5}
  & direct fitting & diagonalization & direct fitting & diagonalization \\
 \hline
 $A^0_3/A^0_1$ & 0.033628(50) & 0.033416(122) & 0.011381(23) & 0.011372(48) \\
 $A^1_{3(5)}/A^1_1$ & 0.03524(231)& 0.03313(47) & 0.011993(728) & 0.011024(142) \\ 
 $R$ & 0.9634(740) & 1.0082(99) & 0.9519(608) & 1.0348(87) \\
\hline 
\end{tabular}
\caption{Results for the direct fit of the correlator matrix in the basis Nll(1), Nnn(2), Hll(3) and for the fit after diagonalizing the correlator matrix in the basis Nll(1), Nln(2), Nnn(3), Nww(4), Hll(5). For the direct fit we state $R^{1,2}_{3,1}$, whereas for the fit after diagonalization $R^{1,2}_{5,1}$.}
 \label{cmfit}
\end{center}
\end{table}

\section{Conclusions}
\label{conclusions}
In the end of this work we want to summarize our results and draw conclusions from them. We found, as do many others working on spectroscopy, that the variational method is a very promising approach since it allows one to clearly separate the individual mass eigenstates of the theory and, furthermore, to find out the ratio of the couplings of a state to different operators included in the basis of the cross correlator matrix. The low cost of simulations in the NRQCD framework also contributes to the success of our analysis. 

We find the quite amazing fact that the ground state and the first few radial excitations of bottomonia have about the same ratio for their couplings to the two local operators we considered. 
The outcome for the radial excitations clearly contradicts the results of \cite{Luo:2005zg}. They cannot find the first radial excitation in their correlator built up from only hybrid currents, not because it is ``skipped,'' but because the coupling between this state and the hybrid operator is too small. With inputs of single symmetric correlator runs we actually are able to conclude that in the correlator built from hybrid operators the amplitude of the radial excitation is down by $O(100)$ compared to the hybrid excitation. The smallness of the coupling is the only reason why one is not able to see any radial excitation in hybrid correlators.

A very important condition for the functionality of our approach is that the ratio of the couplings of the two local operators is independent from the choice of basis. In other words, the local operators have to be orthogonal to the smeared ones used in the basis.
We also show that the $R$ value, on which we mainly base our statements, is rather independent of the scale. However, slight sea quark effects can be observed for $R$.

As a crosscheck for our results we directly fit the correlator matrix without any diagonalization. Although the quality of the fit is comparable for the ground state, it is inferior for the first radial excitation. Separating the individual eigenstates by diagonalization enables us to start the fit much earlier, about $t$ of 5, whereas for the direct fitting the starting time is about $t=12$, leading to larger errors for the radial excitation.

\ack
We wish to thank Andreas Sch\"afer for fostering a large research group, thereby making collaborations like this possible. We would also like to thank the MILC Collaboration for making their configurations publicly available. This work is supported by GSI.

\end{document}